\documentstyle[12pt]{article}
\begin{document}
\font\ninerm = cmr9

\def\footnoterule{\kern-3pt \hrule width \hsize \kern2.5pt}

\pagestyle{empty}
\hfill gr-qc/9603013
\begin{center}
{\large\bf On Local Observations in Quantum Gravity}
\end{center}
\vskip 1.5 cm
\begin{center}
{\bf Giovanni AMELINO-CAMELIA}\\
\end{center}
\begin{center}
{\it Theoretical Physics, University of Oxford,
1 Keble Rd., Oxford OX1 3NP, UK}
\end{center}

\vspace{1cm}
\begin{center}
{\bf ABSTRACT}
\end{center}

{\leftskip=0.6in \rightskip=0.6in

By taking into account both quantum mechanical 
and general relativistic effects,
I derive an equation that describes limitations 
on the measurability of space-time distances
as defined by a material reference system.

}

\vskip 3.4cm

\vfill

\noindent{OUTP-96-09-P 
\hfill February 1996}

\newpage
\baselineskip 12pt plus .5pt minus .5pt
\pagenumbering{arabic}
\pagestyle{plain} 

\section{Introduction}
The construction of a quantum theory incorporating gravity
will certainly require the development of several new "tools",
also to replace some of those we are accustomed to use within
quantum theories in fixed (or classically dynamical) geometry.

Progress in the development of the new tools is possible
even before having a fully consistent quantum gravity;
most notably, 
as seen in the investigation of {\it gedanken} experiments for the
quantum measurement of 
distances[1-5]
and the (semi-classical) quantum analysis of black holes\cite{bh},
one can try to develop intuition for the nature of this tools
by analyzing problems in which
the incompatibility between
quantum mechanics and (classical) general relativity is more evident.

Some of the familiar tools that we need to relinquish 
or modify
have also been identified; in particular, it is clear that
local observables are not easily available 
in a diffeomorphism-invariant quantum theory.
In order to have local observables one can 
introduce a material reference system (MRS) and
include in the analysis
the dynamics of the objects 
that form the reference system\cite{rovelli}.

In this paper I combine the elements of
two of the types of analyses mentioned above.
I consider the quantum measurement of the distances
defined by a MRS.
For the measurement procedure I follow my analysis 
of Ref.\cite{gacmpla}, appropriately adapted to the context
at hand, and as MRS I adopt
one of those discussed by Rovelli in Ref.\cite{rovelli}.
My objective is to provide evidence that the commutation
relations satisfied by the local observables defined by
such MRSs could be very different from the ones
satisfied by the local observables of ordinary
quantum theories in fixed geometry.
This evidence shall be encoded in the structure of a lower
bound on the uncertainty 
present in the quantum measurement under investigation.

\section{Nature of the Uncertainties}
Let me start by reviewing the role of the MRSs.
The difficulties in introducing local observables
in quantum (and even classical) gravity
originate from the fact that diffeomorphism invariance
washes away\cite{rovelli} the individuality of the
points of the universe manifold.
This individuality can be regained by specifying
points by means of some matter, {\it i.e.} introducing
a MRS, but then the analysis has to take
into account the dynamics of this matter.

MRSs have appeared in one form or another
(such as dust, fluids, etc.)
rather frequently in the literature.
I follow Rovelli's discussion\cite{rovelli};
specifically, for simplicity, I consider 
a MRS composed of identical
bodies (defining space points)
with some internal physical variables
defining time instants,
which is one of the MRSs considered in Ref.\cite{rovelli},
and discuss
the quantum measurement
of the distance 
between (the respective centers of mass of) two of such bodies.

As discussed in Refs.\cite{wign,ng,gacmpla}, 
this type of measurement is naturally carried out by exchanging
a light signal between the two bodies.
For conceptual simplicity, let me take one of the two bodies 
to be a clock (after all, a clock is a
body with some internal physical variables
defining time instants).
The measurement could then be performed by ``attaching"  
non-rigidly\footnote{The reader can easily realize 
that rigid attachment is 
not consistent with the causal nature of the theory.} 
a ``light-gun" ({\it i.e.} a device 
capable of sending
a light signal when triggered) and a detector
to the clock,
and ``attaching" a mirror to the other body. 
The system could be set up so that
a light signal be sent toward the mirror
when the clock reads the time $T_i$, and to record
the time $T_f$ shown by the clock when
the light signal is detected by the detector after being 
reflected by the mirror.
Clearly the time $T \! \equiv \! T_f \! - \! T_i$ is related to
the distance $L$; for example,
in Minkowski space and neglecting quantum effects one 
simply finds that
$L = c {T \over 2}$, with $c$ denoting the speed of light.
I am interested in including quantum mechanical and 
general relativistic effects
in the analysis of this measurement
procedure, and therefore
the relation between $T$ 
and $L$ is more complex.
The ``actual" distance $L$ 
and the outcome 
$T$ (time read by the clock) of a measurement procedure,
are related as follows\cite{gacmpla}
\begin{eqnarray}
L = c {T \over 2} \pm \delta L + \Delta L \pm \delta_g L ~,
\label{general}
\end{eqnarray}
where $\delta L$ is the total 
quantum mechanical uncertainty due to 
the quantum mechanical uncertainties in 
the position and velocity of the various
agents in the measurement procedure
(as discussed in
Refs.\cite{wign,ng},
contributions to $\delta L$ come, for example, from the spread 
in the position of the various devices
during the time $T$),
$\Delta L$ is the total 
correction due to the 
gravitational forces among the 
agents in the measurement procedure (for example, one such correction
results from the
gravitational attraction between the photons composing a light signal 
and the devices in the apparatus),
$\delta_g L$ is the total 
quantum mechanical uncertainty
which results from the uncertainties in the
gravitational forces among the 
agents in the measurement procedure (for example, as a result of the 
quantum mechanical spread in the mass of the clock, 
there is an uncertainty in the strength of the
gravitational attraction exerted by the clock on 
the photons composing the light signal).

Concerning $\Delta L$, 
let me point out that, as observed in Ref.\cite{gacmpla},
this contribution to Eq.(\ref{general})  does not play 
any role in my study because
in a context in which both general relativistic and 
quantum mechanical effects are taken into account,
$\Delta L$ represents a correction, not an uncertainty. In fact,
$\Delta L$ can be calculated and taken into account in
the analysis of the outcome of the measurement, therefore leading
to no additional uncertainty.

I intend to investigate
the limitations on the accuracy of quantum measurements of $L$
resulting from the fact that 
it is not possible to render arbitrarily small
the uncertainty introduced in the measurement
of $L$ by the presence of $\delta L$ and $\delta_g L$.

\noindent
$\delta_g L$
has already been considered in Refs.\cite{padma,engl}, 
and it has been found that 
$\delta_g L \geq L_p$, $L_p$ being the Plank length.

\noindent
Concerning $\delta L$
I shall not attempt to derive the absolute lower bound
(whose rigorous derivation 
surely involves an extremely complex analysis) on the
uncertainty of quantum measurements of $L$, 
instead I shall look for a lower bound (which may well be 
quite lower than 
the absolute lower bound) for this uncertainty.
Consistently with this objective, 
the only contribution to $\delta L$ 
that I will consider is $\delta x^{rel}$, defined as 
the uncertainty introduced by the spread 
in the relative position between the center of mass of the clock
and the center of mass of the system composed
by light-gun and detector 
during the time $T$. 
The other contributions to $\delta L$ (given, among several others,
by the spread in the position of the
light-gun, and detector with respect to their center of mass,
and by the spread in the relative position between the
mirror and the second body) 
could obviously only increase 
the lower bound
on the uncertainty that I will present. As done
in Ref.\cite{ng,gacmpla}, I give an intuitive and simple
discussion rather than attempting to find a stronger bound.
By looking for a lower 
bound $min \{ \delta x^{rel} \}$
for $\delta x^{rel}$, I shall get a lower bound for the
uncertainty $\delta L^{tot}$
in the length measurement under consideration, as given by the relation
\begin{eqnarray} 
\delta L^{tot} \equiv \delta L + \delta_g L \geq
\delta x^{rel} + L_p \geq min \{ \delta x^{rel} \} + L_p
~.
\label{defdltot}
\end{eqnarray}
In the following, in order to be able to use spherical symmetry,
I shall also assume 
that the bodies composing the material reference system
(including our clock) are spherical with radius $s$ and
homogeneously-distributed mass $M$.

\section{Derivation of the Bound}
Ignoring for the moment the gravitational effects, 
the evaluation of the spread
in the relative position between 
the center of mass of the clock
and the center of mass of the system composed
by light-gun and detector 
during the time interval $[T_i,T_i+T]$
is a simple quantum mechanical problem.
Following Ref.\cite{wign,gacmpla,ng}, one finds that 
\begin{eqnarray}
\delta x^{rel} \equiv \delta x^{rel}(t \epsilon [T_i,T_i+T]) 
\geq \delta x^{rel}_{t=T_i} 
+ { \hbar T \over \mu \, \delta x^{rel}_{t=T_i} } 
\sim 
\delta x^{rel}_{t=T_i} 
+ { \hbar \over c } 
{ 2 L \over \mu \, \delta x^{rel}_{t=T_i} } 
~,
\label{dawign}
\end{eqnarray}
where $\delta x^{rel}_{t=T_i}$ is 
the initial ({\it i.e.} at the time $t\!=\!T_i$ when the 
light signal is emitted)
spread,
and $\mu$ is the relative mass, related to $M$, the mass of the clock,
and $M_a$, the total mass of the apparatus composed of
light-gun and detector , by the usual relation
\begin{eqnarray}
\mu \equiv {M M_a \over M+ M_a}
~.
\label{mudef}
\end{eqnarray}
(Also note that on the right-hand-side of Eq.(\ref{dawign})
I used the fact that in first
approximation $T \sim 2 L/c$.)

Eq.(\ref{dawign}) can be understood as follows\cite{wign,ng,gacmpla}.
Initially, the wave packet has 
relative-position spread $\delta x^{rel}_{t=T_i}$ and 
relative-velocity spread
$\delta v^{rel}_{t=T_i}$. 
During the time interval $[T_i,T_i+T]$
the uncertainty in the relative position 
is given by
\begin{eqnarray}
\delta x^{rel} \sim \delta x^{rel}_{t=T_i} + \delta v^{rel}_{t=T_i} \, T 
\sim \delta x^{rel}_{t=T_i} + \delta v^{rel}_{t=T_i} \, { 2 L \over c} 
~.
\label{dawign2}
\end{eqnarray}

Eq.(\ref{dawign2}) reproduces Eq.(\ref{dawign}) 
once one takes into account the uncertainty principle, 
which states that
\begin{eqnarray}
\delta x^{rel}_{t=T_i} \, \delta v^{rel}_{t=T_i} \geq {\hbar \over \mu}
~.
\label{up}
\end{eqnarray}

The most important feature of Eq.(\ref{dawign}) is that it indicates that,
with fixed masses $M$ and $M_a$,
there is no way to prepare the $t \! = \! T_i$-wave-packet
so that $\delta x^{rel} \! = \! 0$. In fact, 
Eq.(\ref{dawign}) indicates that
quantum mechanics leads to the following
minimum value of $\delta x^{rel}$ 
\begin{eqnarray}
min \{ \delta x^{rel} \} \sim 
\sqrt{{\hbar T \over \mu}} \sim 
\sqrt{{\hbar \over c} {L\over \mu}}
~,
\label{dminqm}
\end{eqnarray}
where, as I shall continue to do in the following, 
I neglected numerical factors of $O(1)$, which are essentially
irrelevant for the discussion presented in this paper.

Up to this point I have only used quantum mechanics 
and therefore it is not surprising to
discover that, as shown by Eq.(\ref{dminqm}), 
$min \{ \delta x^{rel} \} \! \rightarrow \! 0$ as
$\mu \! \rightarrow \! \infty$. Indeed, 
the uncertainty that I am considering originates from
the uncertainty in the kinematics of the bodies involved
in the measurement procedure,
and clearly this uncertainty vanishes in the limit
of infinite masses (the classical limit) because 
in this limit Eq.(\ref{up}) is consistent 
with $\delta x^{rel}_{t=T_i} \! = \! \delta v^{rel}_{t=T_i} \! = \! 0$.

\noindent
However, the central observation of the present paper is that
this scenario is significantly modified 
when the general relativistic effects 
relevant to our experimental set up are 
taken into account.

It is important to realize that,
with fixed radius $s$ for our spherical clock,
large values of the mass $M$ necessarily lead to great
distorsions of the geometry, and well before 
the $M \! \rightarrow \! \infty$
limit (which is desirable
for reducing the uncertainty given by Eq.(\ref{dminqm}),
since $\mu \! \rightarrow \! \infty$ requires
$M \! \rightarrow \! \infty$) our
measurement procedure can no longer be followed.
In particular, if $M \! \geq \! {\hbar \over c} {s \over L_p^2}$
an {\it horizon} 
forms
around the center of mass of the
clock and it is not 
possible\footnote{If the light-gun is within the clock's horizon
then its light signals cannot reach the mirror.
Its light signals can reach the mirror if instead
the light-gun is outside the clock's horizon; however,
in that case it is still impossible to perform the measurement  
since the light-gun could not ``read" the clock,
and therefore could not be triggered by the 
clock.}
to have a light signal emitted at $t \! = \! T_i$ 
reaching the mirror positioned
on the other body whose distance from the clock is being 
measured; therefore the condition
\begin{eqnarray}
M \leq {\hbar \over c} {s \over L_p^2}
~,
\label{mhor}
\end{eqnarray}
is necessary to perform the measurement.

Since $\mu \! \leq \! M/2$ (see Eq.(\ref{mudef})),
Eqs.(\ref{dminqm}) and (\ref{mhor})
combine to give
\begin{eqnarray}
min \{ \delta x^{rel} \} \sim \sqrt{c T L_p^2 \over s} 
\sim \sqrt{L L_p^2 \over s} 
~.
\label{dmin}
\end{eqnarray}
In turn, this can be combined with Eq.(\ref{defdltot})
to finally obtain a lower bound 
on the uncertainty in the measurement of the distance $L$
\begin{eqnarray}
\delta L^{tot} \geq
min \{ \delta x^{rel} \} + L_p \sim \sqrt{c T L_p^2 \over s} + L_p
\sim \sqrt{L L_p^2 \over s} + L_p
~.
\label{dlmin}
\end{eqnarray}
Notably $min \{ \delta x^{rel} \}$ has introduced an 
$s$- and $L$- dependent
contribution to the lower bound, and actually this contribution
is larger\footnote{Notice that in this context the $s \rightarrow \infty$ 
limit is meaningless,
since it is not possible (not even conceptually) to 
set up the network of bodies necessary to form 
the MRS if each one of the bodies occupies all of space. 
Actually, within a MRS formed by bodies of size $s$ it only makes 
sense to
consider distances $L$ greater than $s$,
and, in order to have as fine as possible a 
network of bodies, one would like $s$ to be small, 
leading to large uncertainty in length measurements.
Moreover, Rovelli's observables are defined only with respect to a given 
MRS, {\it i.e.} in the case I considered they are characterized by 
a specific value of $s$, and therefore increasing $s$ 
one would not reach better and better accuracy in the measurement of 
{\it the same} observable, 
one would instead span a class of {\it different} 
observables which ``live" in different MRSs, 
and are measurable with different accuracy depending on their 
specific value of $s$ (the value of $s$ 
of the MRS they ``live" in).}
than the one previously identified; in fact,
by construction of the MRS here considered,
the distance between the centers of mass of two of the bodies
is always larger than $s$,
so that $min \{ \delta x^{rel} \} \! > \! L_p$.

\section{Discussion and Outlook}
In this paper, 
upon appropriate redefinitions and adaptations,
I succeeded in following,
within the specific framework of one of 
the MRSs considered in Ref.\cite{rovelli},
all the steps of the measurement analysis 
presented in the more general/intuitive
discussion of Ref.\cite{gacmpla}.
This has led to the $s$- and $L$- dependent
lower bound (\ref{dlmin})
on the uncertainty in the measurement of a distance $L$
defined by a MRS characterized by the scale $s$.
This has obvious physical interest on its own,
and can be interpreted as evidence for the need of 
nonconventional commutation relations 
to be enforced on the quantum gravity local observables
defined by MRSs.

There is one viewpoint from which part of this
result is not surprising; in fact, it is nearly always the
case that physics depends nontrivially on all the available 
scales, and therefore, one could have considered 
the possibility that the physics ``seen" by a MRS 
might be dependent on the scales characterizing 
the MRS even without the analysis I presented.
My analysis provides some intuition on how
such dependence might come about.

I also hope that my result
encourages (and gives hints to) future work on
nonconventional commutation relations for 
local observables of quantum gravity.
As a first step toward the introduction of bounds of the 
type (\ref{dlmin}) in candidate quantum gravity theories,
I expect that it could be useful to consider 
the special case $s \sim L_p$
in which Eq.(\ref{dlmin}) takes the form
\begin{eqnarray}
\delta L^{tot} \geq
min \{ \delta x^{rel} \} + L_p \sim \sqrt{c T L_p} + L_p
\sim \sqrt{L L_p} + L_p
~.
\label{dlminsplank}
\end{eqnarray}
Work is in progress investigating the possibility that relations of
the type (\ref{dlminsplank}) might naturally arise in certain
formulations\cite{emn} of non-critical string theory.

A $L$-dependent ({\it i.e.} dependent on the time
required to complete the measurement procedure)
lower bound on the uncertainty 
in the measurement of the distance $L$
could also be important, as observed in Refs.\cite{ng,karo},
in relations to possible mechanisms for (de)coherence
in quantum gravity.
I leave further investigation of this issue for future
studies.

\bigskip
\bigskip
\bigskip
\bigskip
In the preliminary stages of this work I greatly benefited from
stimulating conversations with Carlo Rovelli and John Stachel, 
which I am very happy to acknowledge.

\newpage
\baselineskip 12pt plus .5pt minus .5pt

\end{document}